\documentclass[aps,superscriptaddress,twocolumn,amsmath,amsfonts,amssymb, floatfix]{revtex4-2}

\usepackage[pdftex]{graphicx}
\usepackage{bm,braket}
\usepackage{xcolor}
\usepackage[colorlinks=true,linkcolor=blue,citecolor=blue,urlcolor=blue]{hyperref}
\usepackage{amsmath, amssymb}
\usepackage{mathrsfs}
\usepackage{color}
\usepackage{orcidlink}
\usepackage{ulem}

\begin{document}

\preprint{APS/123-QED}

\title{Enhanced Temperature Sensitivity in Ensemble NV Centers through\\ Improved Optically Detected Magnetic Resonance Spectral Modeling}% Force line breaks with \\

\author{Yuki S. Kato\,\orcidlink{0009-0000-8730-8160}}
 \email{yuki.s.kato@gmail.com}
 \affiliation{
 \mbox{Department of Earth and Space Science, Graduate School of Science, The University of Osaka, Osaka 560-0043, Japan}}%Lines break automatically or can be forced with \\

\author{Shingo Sotoma\,\orcidlink{0000-0001-9528-471X}}%
%\email{shsotoma@kit.ac.jp}
\affiliation{%
\mbox{Faculty of Molecular Chemistry and Engineering, Kyoto Institute of Technology, Sakyo-ku, Kyoto 606-8585, Japan}
}%
 
\author{Keisuke Fujita\,\orcidlink{0000-0002-1900-5325}}%
\affiliation{%
\mbox{Premium Research Institute for Human Metaverse Medicine, The University of Osaka, Osaka 565-0871, Japan}
}%

\author{Masanori Fujiwara\,\orcidlink{0000-0002-6801-0905}}%
%\email{shsotoma@kit.ac.jp}
\affiliation{%
Institute for Chemical Research, Kyoto University, Uji, Kyoto 611-0011, Japan
}%

\author{\\Izuru Ohki\,\orcidlink{0000-0002-0667-5936}}%
\affiliation{%
Institute for Chemical Research, Kyoto University, Uji, Kyoto 611-0011, Japan
}%

\author{Yuichiro Matsuzaki\,\orcidlink{0000-0001-6814-3778}}%
%\email{shsotoma@kit.ac.jp}
\affiliation{%
Department of Electrical, Electronic, and Communication Engineering, Faculty of Science and Engineering, Chuo University, Tokyo, Japan
}%

\author{Norikazu Mizuochi\,\orcidlink{0000-0003-3099-3210}}%
%\email{shsotoma@kit.ac.jp}
\affiliation{%
Institute for Chemical Research, Kyoto University, Uji, Kyoto 611-0011, Japan
}%

\author{Yoshie Harada\,\orcidlink{0000-0003-2249-5553}}%
\email{yharada@protein.osaka-u.ac.jp}
\affiliation{%
\mbox{Premium Research Institute for Human Metaverse Medicine, The University of Osaka, Osaka 565-0871, Japan}
}%
\affiliation{%
\mbox{Center for Quantum Information and Quantum Biology, The
University of Osaka, Osaka 560-0043, Japan}
}%

\date{\today}% It is always \today, today,

\begin{abstract}
Nitrogen-vacancy (NV) center ensembles provide a powerful platform for high-precision temperature sensing, with ongoing efforts to further enhance their measurement performance. In ensemble NV optically detected magnetic resonance (ODMR) spectra, commonly used Lorentzian and Voigt fitting models fail to accurately describe the spectral shape near the resonance frequency, leading to degraded precision in resonance-frequency determination and, consequently, temperature estimation. In this work, we analytically establish a new fitting method, termed dip–peak fitting, for extracting the resonance frequency from ensemble cw-ODMR spectra. Starting from a physical model that describes ensemble cw-ODMR spectra as a convolution of single-NV responses with distributed zero-field splitting and strain, we show that the spectral feature near resonance can be accurately approximated by a single Lorentzian function with a background term. The proposed fitting model reproduces the cw-ODMR spectrum around resonance more faithfully than conventional approaches, enabling faster and more accurate resonance-frequency determination under weaker microwave excitation. Experiments using fluorescent nanodiamond ensembles confirm the robustness and applicability of this method for high-precision temperature sensing.
\end{abstract}

\maketitle

%\tableofcontents

\section{\label{sec:level1}INTRODUCTION}
The negatively charged nitrogen-vacancy (NV) center in diamond is a versatile solid-state quantum sensor, as its spin state can be optically initialized, coherently manipulated, and read out under ambient conditions at room temperature~\cite{gruber_scanning_1997,doherty_nitrogen-vacancy_2013,schirhagl_nitrogen-vacancy_2014}.
For many sensing applications, achieving high sensitivity requires sufficient photon collection and signal stability, which can be challenging for measurements based on single NV centers.
Ensembles of NV centers address this limitation by providing enhanced signal-to-noise ratios through collective optical readout, enabling high-sensitivity measurements under practical experimental conditions~\cite{hayashi_optimization_2018}. Fluorescent nanodiamonds (FNDs) hosting NV ensembles combine this ensemble signal enhancement with nanoscale probe dimensions, allowing quantum sensing with both high sensitivity and nanoscale spatial resolution~\cite{schirhagl_nitrogen-vacancy_2014}.
These properties enable detection of various physical quantities, including magnetic fields~\cite{balasubramanian_nanoscale_2008,maze_nanoscale_2008}, electric fields~\cite{dolde_electric-field_2011,dolde_nanoscale_2014}, and temperature~\cite{kucsko_nanometre-scale_2013}. 

\begin{figure}[b]
\includegraphics[width=\columnwidth]{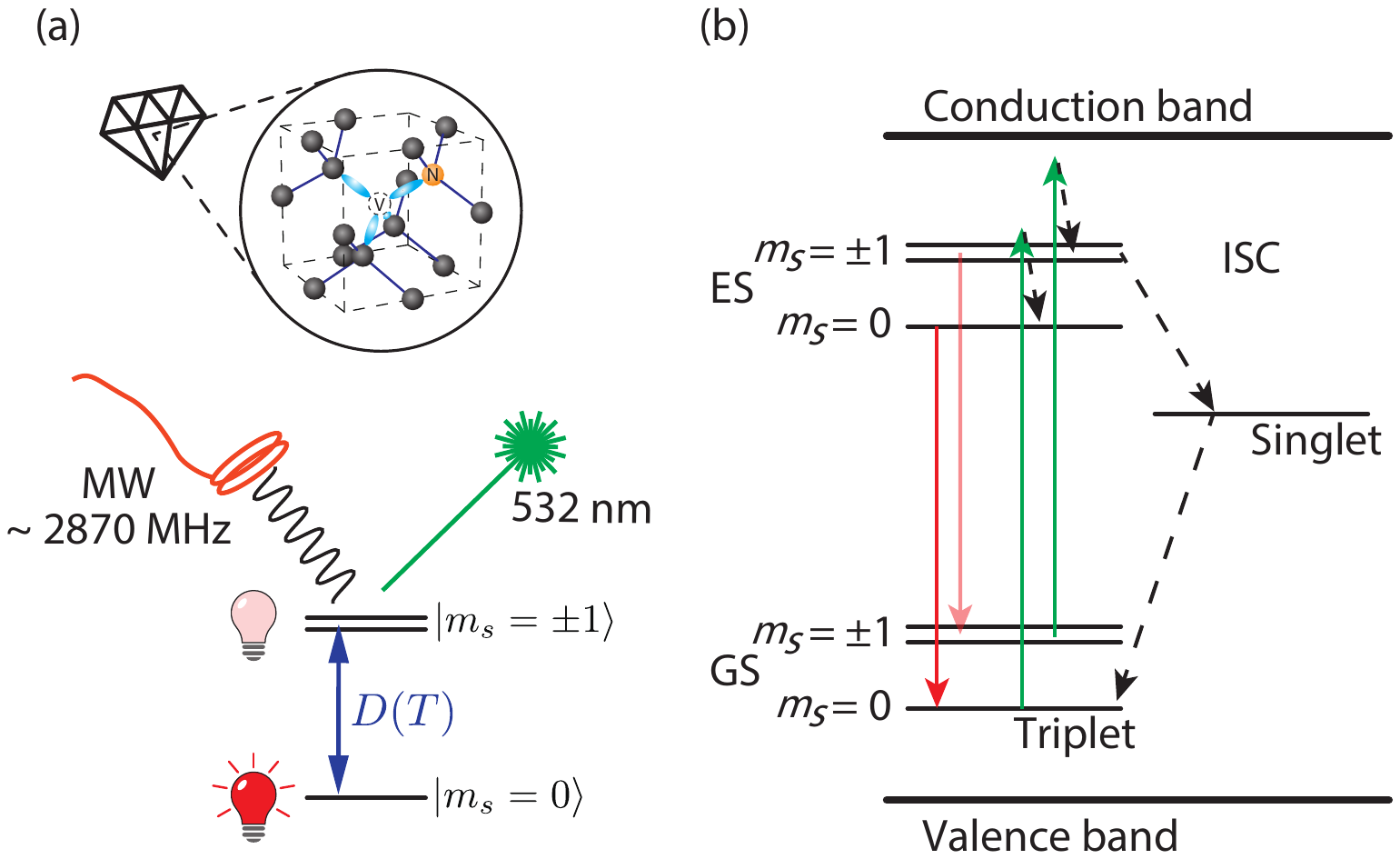}% Here is how to import EPS art
\caption{\label{fig:INT}(a) Schematic illustration of a nitrogen–vacancy center (NVC) in diamond. The NVC is a point defect in the diamond lattice, consisting of a substitutional nitrogen atom adjacent to a lattice vacancy. The NVC is a spin-1 system and exhibits unique optical properties.
(b) Energy-level diagram of a NVC.}
\end{figure}

The NV center has a spin-1 ground state, with the
$\ket{m_s = \pm1}$  levels separated from the $\ket{m_s = 0}$  level by
the zero-field splitting of approximately 2.87 GHz [Fig.~\ref{fig:INT}(a)].
Owing to spin-state-dependent fluorescence under green
laser excitation, the NV electron spin can be efficiently
initialized and read out optically\cite{gali_ab_2019}[Fig.~\ref{fig:INT}(b)].
Sweeping the microwave frequency around the zero-field
splitting gives rise to optically detected magnetic resonance
(ODMR) spectra.
Because the zero-field splitting depends on temperature via thermal expansion of the diamond lattice, shifts in the ODMR spectra allow for quantitative thermometry
\cite{acosta_temperature_2010,cambria_physically_2023}.

Accurate temperature sensing is crucial in a wide range of physical and biological systems.
Compared with other fluorescent thermometers, FNDs enable temperature to be extracted independently of fluctuations in other physical parameters and chemical environments, thereby allowing for reliable temperature measurements\cite{sekiguchi_fluorescent_2018}.
This capability has led to their widespread use in biological systems, including intracellular thermometry\cite{kucsko_nanometre-scale_2013,sotoma_situ_2021,chuma_implication_2024,wu_singlecell_2026,lee_organelle-specific_2025,so_small_2024}.

There are two principal strategies for temperature measurement using FNDs. One approach tracks temperature changes within the same diamond, while the other compares temperatures across different diamonds at different spatial and temporal locations. In the former case, temperature variations can be inferred by monitoring fluorescence intensity at one or several fixed microwave frequencies\cite{kucsko_nanometre-scale_2013,wu_singlecell_2026,fujiwara_real-time_2020,so_small_2024}. In contrast, the latter approach requires sweeping the microwave frequency to acquire ODMR spectra, from which the resonance frequency is identified via spectral fitting\cite{sotoma_situ_2021,chuma_implication_2024,lee_organelle-specific_2025}.

To date, Lorentzian and Voigt functions have been widely employed to fit ensemble ODMR spectra. The Lorentzian function is based on the response of an ideal two-level system and is typically expressed as a superposition of two Lorentzian components corresponding to the $\ket{m_s = 0} \rightarrow (1/\sqrt{2})\left(\Ket{1}-\Ket{-1}\right)$ and $\ket{m_s = 0} \rightarrow (1/\sqrt{2})\left(\Ket{1}+\Ket{-1}\right)$ transitions. In practice, however, each NV center experiences a distinct local environment, and consequently, not all NV centers exhibit identical spectral line shapes. As a result, Lorentzian functions alone fail to adequately describe ensemble ODMR spectra\cite{kubo_strong_2010,yamamoto_nanodiamond_2025}. To address this limitation, Voigt functions—assuming a distribution of lattice strain or electric-field interactions among NV centers—have also been employed. Nevertheless, even Voigt functions cannot fully reproduce ensemble ODMR spectra. 

Such fitting functions, which do not fully capture the ODMR spectrum, have limited the achievable performance of temperature measurements.
In practice, the achieved temperature precision has been substantially worse than the limits expected from fundamental noise sources, such as shot-noise-limited detection.
This discrepancy indicates that further improvements require the development of more appropriate fitting models.
To address this issue, various approaches and fitting strategies have been proposed and investigated in previous studies\cite{hayashi_optimization_2018,yamamoto_nanodiamond_2025}. In this context, alternative fitting strategies have been explored that focus on spectral features in the immediate vicinity of the resonance frequency, rather than on the global ODMR line shape.

Kołodziej et al. proposed a practical fitting approach in which the peak-like feature within the cw-ODMR dip in the vicinity of the resonance frequency is modeled by a single Lorentzian function plus a background term\cite{kolodziej_multimodal_2024}. However, the physical basis of this fitting form and its quantitative utility have not been clarified, and the method has not yet been widely adopted. In this paper, we analytically elucidate the physical origin of this fitting model—hereafter referred to as dip-peak fitting—and rigorously demonstrate its effectiveness.

We begin by expressing the ensemble cw-ODMR spectrum as a convolution integral of single-NV cw-ODMR spectra. Through an analytical transformation of this expression, we show that the spectral feature in the vicinity of the resonance frequency can be described by a single Lorentzian function. Experimentally, we demonstrate that dip-peak fitting reproduces the cw-ODMR spectrum near resonance more faithfully than conventional fitting methods and enables high-precision estimation of the resonance frequency with significantly reduced acquisition time. We further investigate the range of microwave excitation strengths compatible with dip–peak fitting, and demonstrate that accurate fitting is achievable over a wide range, with optimal performance obtained at relatively low excitation powers. Finally, by varying the temperature of FNDs, we confirm that dip-peak fitting captures the temperature-induced resonance shifts observed using established fitting approaches.

\section{\label{sec:level1}EXPERIMENTAL METHOD}

\begin{figure}[b]
\includegraphics[width=\columnwidth]{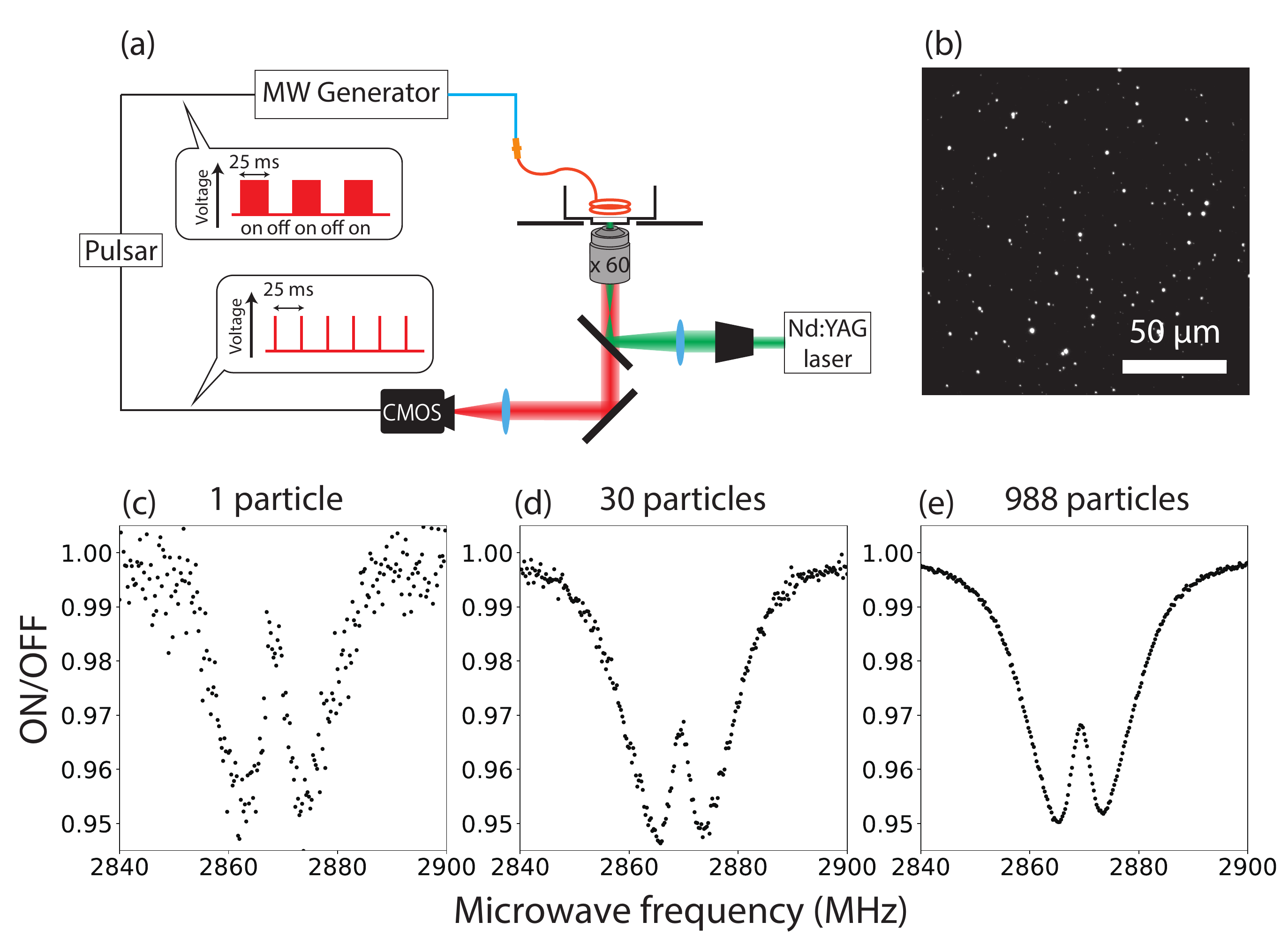}% Here is how to import EPS art
\caption{\label{fig:EXP} (a) Optical setup for cw-ODMR measurements.
(b) Fluorescence image of bright spots from FNDs dispersed on a glass substrate.
(c–e) cw-ODMR spectra obtained from ensembles consisting of a single bright spot (c), 30 bright spots (d), and 988 bright spots (e).
}
\end{figure}
Nanodiamonds with an average diameter of approximately 100 nm, synthesized using the high-pressure high-temperature (HPHT) method, were used. To generate optically active NVCs, we introduced vacancies via electron irradiation with 2 MeV at a fluence of $4 \times 10^{18} \text{cm}^{-2}$, followed by thermal annealing to pair nitrogen atoms and vacancies. Subsequent air oxidation removed sp$^2$ carbon from the particle surface, yielding FNDs. 

The FNDs were spin-coated onto a glass substrate, yielding a uniform dispersion dominated by isolated single particles [Fig.~\ref{fig:EXP}(b)]. The cw-ODMR spectrum measurements were performed using a homemade microscope [Fig.~\ref{fig:EXP}(a)]. A continuous neodymium-doped yttrium aluminum garnet (Nd:YAG) laser at 532 nm illuminated FNDs to initialize and read out the spin state of NVCs on an inverted microscope system. Fluorescence from the NVCs was collected by an oil immersion objective lens, spectrally filtered by a long-pass filter to suppress the excitation light, and imaged onto a camera. 

Microwave (MW) excitation was applied using a two-turn copper coil with a diameter of approximately 1 mm, positioned directly above the coverslip to irradiate the sample at frequencies near the electron spin resonance of the NVCs. The MW signal generated by a microwave source was amplified using linear microwave power amplifiers and delivered to the coil via a coaxial cable. 

cw-ODMR spectra were obtained from the ratio of fluorescence intensities measured with the MW field switched on and off. While the cw-ODMR spectrum acquired from single bright spots exhibits significant noise, summing the fluorescence signals from multiple bright spots yields a cw-ODMR spectrum with substantially reduced noise [Fig.~\ref{fig:EXP}(c, d, e)].

\section{\label{sec:level1}MODEL}
We propose a novel fitting method for cw-ODMR spectra. When the external magnetic field and nuclear spins can be neglected, The Hamiltonian of the NV center is described as
\begin{eqnarray}
H = D(T) S_z^2 +  E_1\left( S_x^2 - S_y^2 \right)
+  E_2 \left( S_x S_y + S_y S_x \right)
 \label{eq:one}.
\end{eqnarray}
where $S_i$ ($i=x,y,z$) are the spin-1 operators of the NV
electron spin, $D(T)$ is the temperature-dependent
zero-field splitting parameter, and $E_1$ and $E_2$
describe transverse strain- or electric-field-induced
splittings\cite{hayashi_optimization_2018}. The eigenvalues of this Hamiltonian are given by
\begin{eqnarray}
   0 , \qquad
    D - \sqrt{E_1^{2} + E_2^{2}}, \qquad
     D + \sqrt{E_1^{2} + E_2^{2}}
\end{eqnarray}
Assuming that the cw-ODMR spectrum of a single NV center can be described as a linear superposition of Lorentzian functions centered at $D - \sqrt{E_1^{2} + E_2^{2}}$ and $D + \sqrt{E_1^{2} + E_2^{2}}$, it can be written as
\begin{eqnarray}
P_1(\omega) &=& 1 -\lambda'\Bigg[L\left(\omega, D+\sqrt{E_1^2+E_2^2}, \Gamma \right)\\&&+L\left(\omega,D-\sqrt{E_1^2+E_2^2}, \Gamma \right)\Bigg]
\end{eqnarray}
where $L\left(x, x_0, \gamma\right)=\gamma / [ \pi \{ (x - x_0)^2 + \gamma^2 \} ]$ is a Lorentzian function, and $\lambda'$ is a scaling factor corresponding to the ODMR contrast.
The ensemble cw-ODMR spectrum is formed by the superposition of single-NV cw-ODMR spectra. Monte Carlo simulations suggest that the sharp cw-ODMR spectrum of each individual NV center varies due to fluctuations in the parameters $D$, $E_1$, and $E_2$, which follow Lorentzian distributions (see Appendix A). Consequently, in systems containing a sufficiently large number of NV centers, the ensemble cw-ODMR spectrum $P$ can be described as
\begin{widetext}
\begin{eqnarray}
P &=&1 - \lambda'\int_{-\infty}^{\infty} dD \int_{-\infty}^{\infty} d E_1 \int_{-\infty}^{\infty} d E_2 L\left(E_1, 0, \gamma_{E_1}\right) L\left(E_2, 0, \gamma_{E_2}\right) L\left(D, D_0, \gamma_D\right)
\\ &&\qquad\qquad\times\left[L\left(\omega, D+\sqrt{E_1^2+E_2^2}, \Gamma \right)+L\left(\omega,D-\sqrt{E_1^2+E_2^2}, \Gamma \right)\right]
\end{eqnarray}
\end{widetext}
Here, for simplicity, we set $\gamma_{E_1} = \gamma_{E_2} \equiv \gamma_{E}$. This assumption reflects the comparable statistical spreads of $E_1$ and $E_2$ and is justified by numerical simulations. For the integration over $D$ the convolution of two Lorentzian functions yields another Lorentzian function whose center and half width at half maximum are given by the sums of those of the original functions. By introducing $r = \sqrt{E_1^2+E_2^2}$, the integration can be carried out explicitly as
\begin{widetext}
\begin{eqnarray}
P 
&=&1 - \lambda'\int_0^\infty d r \frac{4 \gamma_E r}{\pi \sqrt{\gamma_E^2+r^2}\left(r^2+2 \gamma_E^2\right)}[L(\omega - D_0,r, \Gamma')+L(\omega - D_0,-r,  \Gamma')]\\ 
&=&1 - \lambda'\left(L(\omega,D_0,\sqrt2\,\gamma_E+\Gamma')-
L(\omega,D_0,\sqrt2\,\gamma_E-\Gamma')
+ \frac{8\Gamma'}{\pi^2}\Re\left[ \frac{\beta}{\sqrt{u_+}}
\ln\!\left(\frac{1+\sqrt{u_+}}{1-\sqrt{u_+}}\right) \right]\right)
\label{eq:analytic_solution}
\end{eqnarray}
\end{widetext}
where $\Gamma' = \Gamma + \gamma_D$, and
\begin{eqnarray}
u_+ &=&
\frac{(\omega - D_0 + i \Gamma')^2 + \gamma_E^2}{\gamma_E^2},
\end{eqnarray}
\begin{eqnarray}
\beta &=&
\frac{\omega - D_0 + i \Gamma'}
{2 i \Gamma' \left((\omega - D_0)^2 + 2\gamma_E^2 - \Gamma'^2 + 2 i (\omega - D_0) \Gamma'\right)}.
\end{eqnarray}
The results of this calculation are presented in Appendix B. These results demonstrate that the emergence of a small peak at the bottom of the dip in the ensemble cw-ODMR spectrum, as observed in [Fig.~\ref{fig:EXP}], originates from lattice-strain fluctuations whose magnitude exceeds the half width at half maximum of the single-NV cw-ODMR spectrum. This interpretation is consistent with previous reports on high-density NV ensembles\cite{matsuzaki_optically_2016,mittiga_imaging_2018}.

In the vicinity of this small peak, namely around 
$\omega = D_0$ the ensemble cw-ODMR spectrum can be approximated as
\begin{equation}
P(\omega)
= 
A + B\,L(\omega,D_0,\Gamma_{\mathrm{dip}}) + o\bigl((\omega - D_0)^6\bigr)
\label{eq:dip_peak}
\end{equation}
Here, the parameters are determined by matching the expansion of
$P(\omega)$ up to fourth order at $\omega=D_0$:
\begin{align}
\Gamma_{\mathrm{dip}}
&=
\sqrt{
-12\frac{P''(D_0)}{P^{(4)}(D_0)}
},\\
B
&=
-\frac{\pi}{2}\Gamma_{\mathrm{dip}}^3 P''(D_0),\\
A
&=
P(D_0)+\frac{\Gamma_{\mathrm{dip}}^2}{2}P''(D_0).
\end{align}
Plots of this approximation are presented in Appendix~B.
For the parameter range
$0.2\le \Gamma'/\gamma_E \le0.6$,
the Lorentzian approximation reproduces the central dip--peak structure
with
$R^2>0.995$
within the interval bounded by the two inflection points of the 
ensemble cw-ODMR spectrum.

These calculations and approximations provide a new fitting framework for a cw-ODMR spectrum exhibiting a small peak at the bottom of the dip. We refer to this approach—in which the central peak feature embedded within the cw-ODMR dip is modeled using a single Lorentzian term $B\,L(\omega,D_0,\Gamma_{\mathrm{dip}})$ combined with a background term \(A\) [Eq.~\eqref{eq:dip_peak}]—as the dip–peak fitting.

\section{\label{sec:level1}RESULTS AND DISCUSSION}
\subsection{\label{sec:level2}Residual comparison between conventional and dip–peak fitting}

\begin{figure}[t]
\includegraphics[width=0.95\columnwidth]{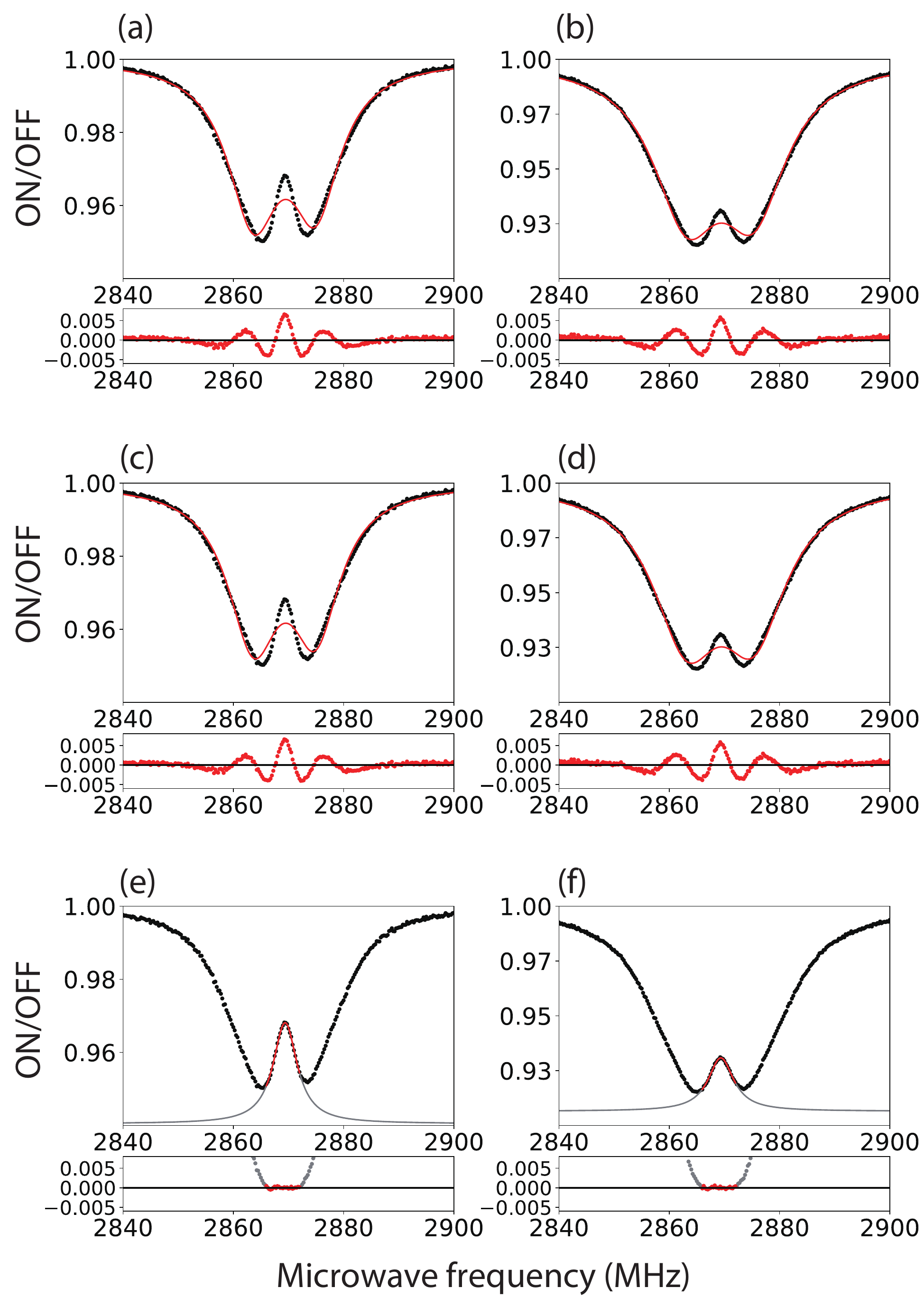}% Here is how to import EPS art
\caption{\label{fig:fig3} Fitting results of cw-ODMR spectra. cw-ODMR spectra measured under weak microwave excitation (a, c, e) and strong microwave excitation (b, d, f) are fitted using a Lorentzian function (a, b), a Voigt function (c, d), and a dip–peak fitting model (e, f). Black dots represent the experimental cw-ODMR data, while red solid lines indicate the fitting results. For panels (e) and (f), the fitting results within the fitting range are shown as red solid lines, while those outside the range are shown as gray solid lines. Red dots shown below each spectrum correspond to the residuals between the experimental data and the fitting curves. In panels (e) and (f), the residuals within the fitting range are shown as red dots, while those outside the range are shown as gray dots.}
\end{figure}

Fig.~\ref{fig:fig3} shows ensemble cw-ODMR spectra with exceptionally low noise, obtained by averaging over approximately 1000 bright spots under conditions of strong and weak microwave excitation. The spectra measured were fitted using conventional Lorentzian fitting, Voigt fitting, and the dip–peak fitting method newly introduced in this work. For the conventional methods, the full spectral range was used, whereas the dip–peak fitting was applied only within the frequency range of 2866–2872 MHz. In the dip–peak fitting, the center frequency, linewidth, amplitude, and offset of the Lorentzian function were treated as fitting parameters.

The conventional Lorentzian and Voigt fittings exhibit large residuals in the vicinity of the resonance frequency, indicating that these models do not adequately describe the cw-ODMR spectra in this region[Fig.~\ref{fig:fig3}(a,b,d,e)]. In contrast, the dip–peak fitting reproduces the spectral features remarkably well in both excitation conditions, particularly capturing the small peak embedded within the resonance dip[Fig.~\ref{fig:fig3}(c,f)].
Quantitatively, within the frequency range of 2866–2872 MHz around the resonance, 
the coefficient of determination $R^2$ obtained using conventional Lorentzian and Voigt fittings falls below 0.5, 
whereas the present model yields $R^2$ values exceeding 0.99 under both excitation conditions.

These results demonstrate that the dip–peak fitting provides a more accurate description of the cw-ODMR spectrum near the resonance frequency than the previously employed fitting approaches.

\subsection{\label{sec:level2}Resonance Frequency Determination Capability and the Law of Large Numbers}

To demonstrate that the dip–peak fitting is suitable for determining the resonance frequency more reliably than conventional fitting methods, we analyzed cw-ODMR spectra with different noise levels by varying the number of bright spots included in the ensemble averaging. Since each field of view contains approximately 1000 bright spots, we randomly selected $n$ bright spots from the field of view to construct an ensemble cw-ODMR spectrum. This spectrum was then fitted using each fitting method, and the resonance frequency was extracted. By repeating this procedure multiple times, we evaluated the statistical variation of the resonance frequency determined from ensemble cw-ODMR spectra composed of $n$ bright spots.

\begin{figure}[b]
\includegraphics[width=\columnwidth]{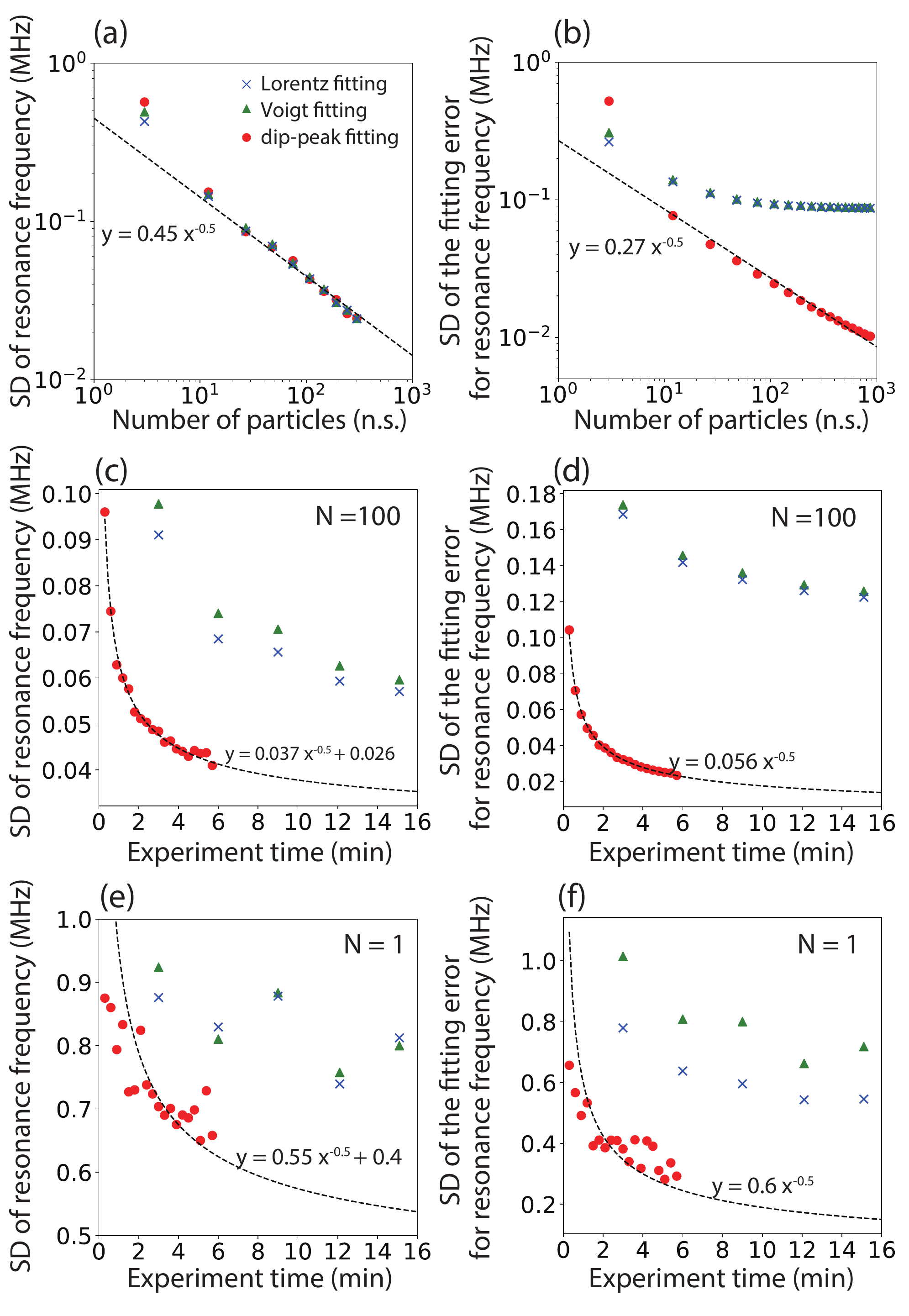}% Here is how to import EPS art
\caption{\label{fig:fig4} Dependence of the resonance-frequency fluctuation and fitting uncertainty on the number of bright spots and the measurement time. When the number of bright spots constituting the cw-ODMR spectrum is increased (a, b) and when the measurement time is increased (c, d, e, f) the resonance-frequency fluctuation (a, c, e) and the fitting error (b, d, f) are evaluated for different fitting methods. Blue crosses indicate results obtained using Lorentzian fitting, green triangles correspond to Voigt fitting, and red circles represent results obtained using the dip–peak fitting model.
The effective measurement times for panels (a, b) are 15 min for Lorentzian and Voigt fitting, and 1.5 min for the dip–peak fitting model. Panels (c, d) are based on ensemble cw-ODMR spectra composed of 100 bright spots, while panels (e, f) use cw-ODMR spectra from a single bright spot. The dashed lines represent fits to the resonance-frequency fluctuations obtained using the dip–peak fitting model, assuming a scaling proportional to the inverse square root of the number of bright spots or the measurement time. The corresponding fitting parameters are indicated in the figure.}
\end{figure}

As the number of bright spots was increased, the variance of the resonance frequencies obtained from all fitting methods decreased in accordance with the law of large numbers[Fig.~\ref{fig:fig4}(a)]. The dip–peak fitting achieved a precision in determining the resonance frequency that is comparable to that of the conventional fitting approaches.

We also evaluated the fitting uncertainty estimated from the diagonal elements of the covariance matrix. For the conventional fitting methods, the fitting functions do not adequately describe the cw-ODMR spectra, and as a result, the estimated fitting uncertainties do not follow the law of large numbers even as the number of averaged bright spots is increased[Fig.~\ref{fig:fig4}(b)]. 
In particular, the fitting uncertainty does not decrease below approximately 0.1 MHz even for the largest number of averaged bright spots considered. 
Moreover, the fitting uncertainties are significantly larger than the actual statistical variations of the resonance frequencies obtained above.

In contrast, for the dip–peak fitting, the fitting function accurately captures the spectral features of the cw-ODMR spectra, leading to fitting uncertainties that decrease in accordance with the law of large numbers. The estimated fitting uncertainty is comparable to the empirical standard deviation of the extracted resonance frequencies, amounting to approximately 0.6 times the observed statistical variation.  This property is particularly advantageous for practical temperature measurements, as it enables a reliable estimation of the confidence in the determined resonance frequency directly from the fitting uncertainty.

Next, we examined, using the same procedure, how the statistical variation of the resonance frequency and the fitting uncertainty depend on the measurement time for ensemble cw-ODMR spectra composed of 100 bright spots and a single bright spot[Fig.~\ref{fig:fig4}(c,d,e,f)]. For both conditions, the variance of the resonance frequency decreased with increasing measurement time, consistent with the law of large numbers. At a measurement time of 6 min, the empirical standard deviation of the resonance frequency is reduced to approximately 0.6 times its initial value for spectra composed of 100 bright spots and to approximately 0.75 times for a single bright spot. 
The dip–peak fitting therefore enables a more precise determination of the resonance frequency within a shorter measurement time compared to conventional fitting methods. 
This improvement is partly attributable to the fact that the microwave frequency sweep range required for the dip–peak fitting is approximately one tenth of that required for the conventional approaches. When the cw-ODMR spectrum of a single bright spot was analyzed using the dip–peak fitting, a residual uncertainty of approximately 0.4 MHz remained even for long measurement times[Fig.~\ref{fig:fig4}(e)]. This residual error is attributed to the intrinsic variation of the resonance frequency inherent to the particle itself. 

With respect to the fitting uncertainty, the dip–peak fitting follows the law of large numbers as the measurement time increases, whereas conventional fitting methods do not exhibit such behavior.

A consistent trend was observed across all investigated FND samples, including those with different NV center concentrations and from different manufacturers (see Appendix C). These results demonstrate that the dip–peak fitting enables the resonance frequency to be determined with higher accuracy within a shorter measurement time than the conventional fitting methods. In addition, it allows for a more reliable estimation of the confidence in the extracted values based on the fitting uncertainty.

\subsection{\label{sec:level2}Appropriate microwave intensity for dip–peak fitting}
\begin{figure}[b]
\includegraphics[width=\columnwidth]{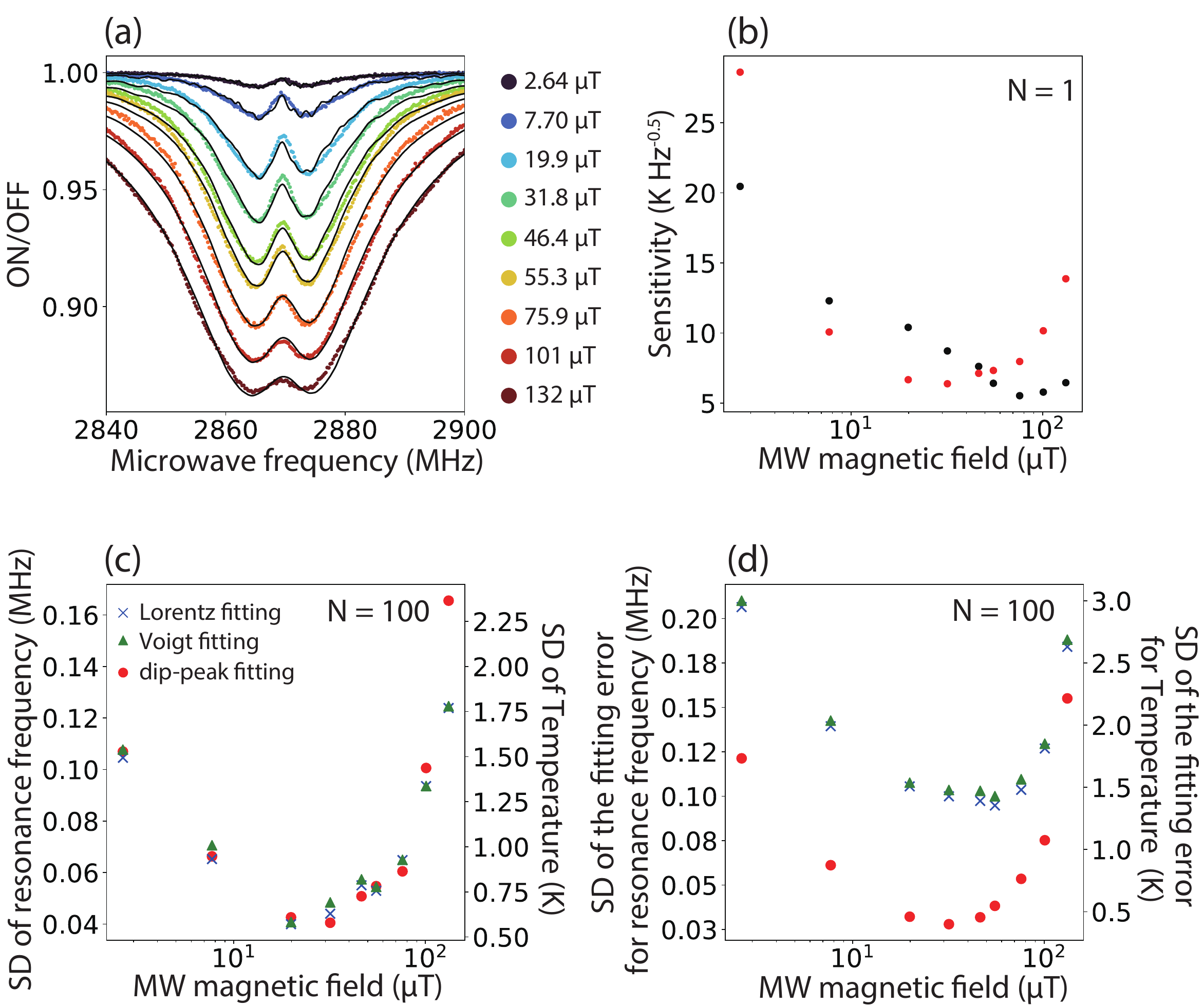}% Here is how to import EPS art
\caption{\label{fig:fig5}(a) cw-ODMR spectra measured at different microwave powers. Black solid lines represent results of Monte Carlo simulations (Appendix A). The microwave power was estimated from the Rabi frequency obtained by fitting the Monte Carlo simulations.
(b) Microwave-power dependence of the shot-noise-limited sensitivity obtained using Lorentzian fitting, Voigt fitting (black dots), and the dip–peak fitting model (red dots).
(c, d) Microwave-power dependence of the resonance-frequency fluctuation (c) and the fitting accuracy (d) for each fitting model. Blue crosses indicate results obtained using Lorentzian fitting, green triangles correspond to Voigt fitting, and red circles represent results obtained using the dip–peak fitting model. The effective measurement times for panels (c, d) are 15 min for Lorentzian and Voigt fitting, and 1.5 min for the dip–peak fitting model.}
\end{figure}
To determine the microwave power suitable for temperature measurements when using the dip–peak fitting, we investigated the dependence on microwave power using nine cw-ODMR spectra measured at different microwave powers[Fig.~\ref{fig:fig5}(a)]. The shot-noise-limited sensitivity is given by the following expression, as
\begin{equation}
    \mu = \frac{1}{(\text{slope})_\text{max}c_T\sqrt{N}}
\end{equation}
Where $(\text{slope})_\text{max}$ is the maximum slope of the cw-ODMR spectrum and $c_T$ is the temperature dependence of the zero-field splitting. We calculated the shot-noise-limited sensitivity for both the conventional fitting methods and the dip–peak fitting. For the conventional fitting, the maximum slope of the spectrum in the frequency range from 2840 MHz to 2865 MHz was used, whereas for the dip–peak fitting, the maximum slope in the frequency range from 2865 MHz to 2870 MHz was employed.

The minimum shot-noise-limited sensitivity was found
to be comparable between the conventional methods and
the dip–peak fitting, with a minimum value of
approximately 5 $\mathrm{K}/\sqrt{\mathrm{Hz}}$ for a single particle[Fig.~\ref{fig:fig5}(b)]. However, the microwave power corresponding to the minimum sensitivity differed between the two approaches. For the dip–peak fitting, the minimum sensitivity was obtained at a microwave power corresponding to a cw-ODMR contrast of approximately 5\% (ON/OFF), whereas for the conventional methods it occurred at approximately 10\%. These results suggest that, under low microwave power conditions, the dip–peak fitting has the potential to achieve higher measurement precision than the conventional fitting approaches.

Using the same procedure as described in the previous section, we experimentally evaluated the statistical variation of the resonance frequency and the fitting precision using ensemble cw-ODMR spectra composed of 100 bright spots[Fig.~\ref{fig:fig5}(c,d)]. The results indicate that high-precision determination of the resonance frequency is maintained over a wide range of microwave excitation strengths, with the highest precision achieved when the maximum cw-ODMR contrast is approximately 5\% (ON/OFF), which is consistent with the trend predicted from the shot-noise-limited sensitivity analysis.

To further quantify the performance, we compared the experimentally observed temperature variation with the shot-noise-limited uncertainty. 
For the conventional fitting method, although the effective acquisition time is approximately 900 s, the experimentally observed temperature variation is about 25 times larger than the shot-noise limit. This indicates that the measurement operates far from the fundamental noise limit.
In contrast, for the dip–peak fitting, with an effective acquisition time of only 90 s, the experimentally observed variation is approximately 10 times larger than the shot-noise limit. Thus, the dip–peak fitting operates substantially closer to the shot-noise-limited regime despite the shorter acquisition time.

These results demonstrate that, when using the dip–peak fitting, the optimal measurement condition corresponds to a maximum cw-ODMR contrast of approximately 5\% (ON/OFF). 
This value is lower than the contrast of approximately 10\% predicted from the shot-noise limit as optimal when using conventional fitting methods, indicating that experiments employing dip–peak fitting can be performed at lower microwave power.
Because reducing the microwave power significantly suppresses heating from the microwave coil, the dip–peak fitting provides a highly advantageous approach for practical temperature measurements. 
Furthermore, comparison with the shot-noise-limited sensitivity reveals that the experimentally observed temperature variation using dip–peak fitting is approximately ten times the shot-noise limit, whereas the conventional approach remains about twenty-five times above this fundamental bound. These results demonstrate that dip–peak fitting enables operation significantly closer to the shot-noise-limited regime while reducing  microwave power.

\subsection{\label{sec:level2}Temperature dependence of resonance frequency}

\begin{figure}[t]
\includegraphics[width=0.95\columnwidth]{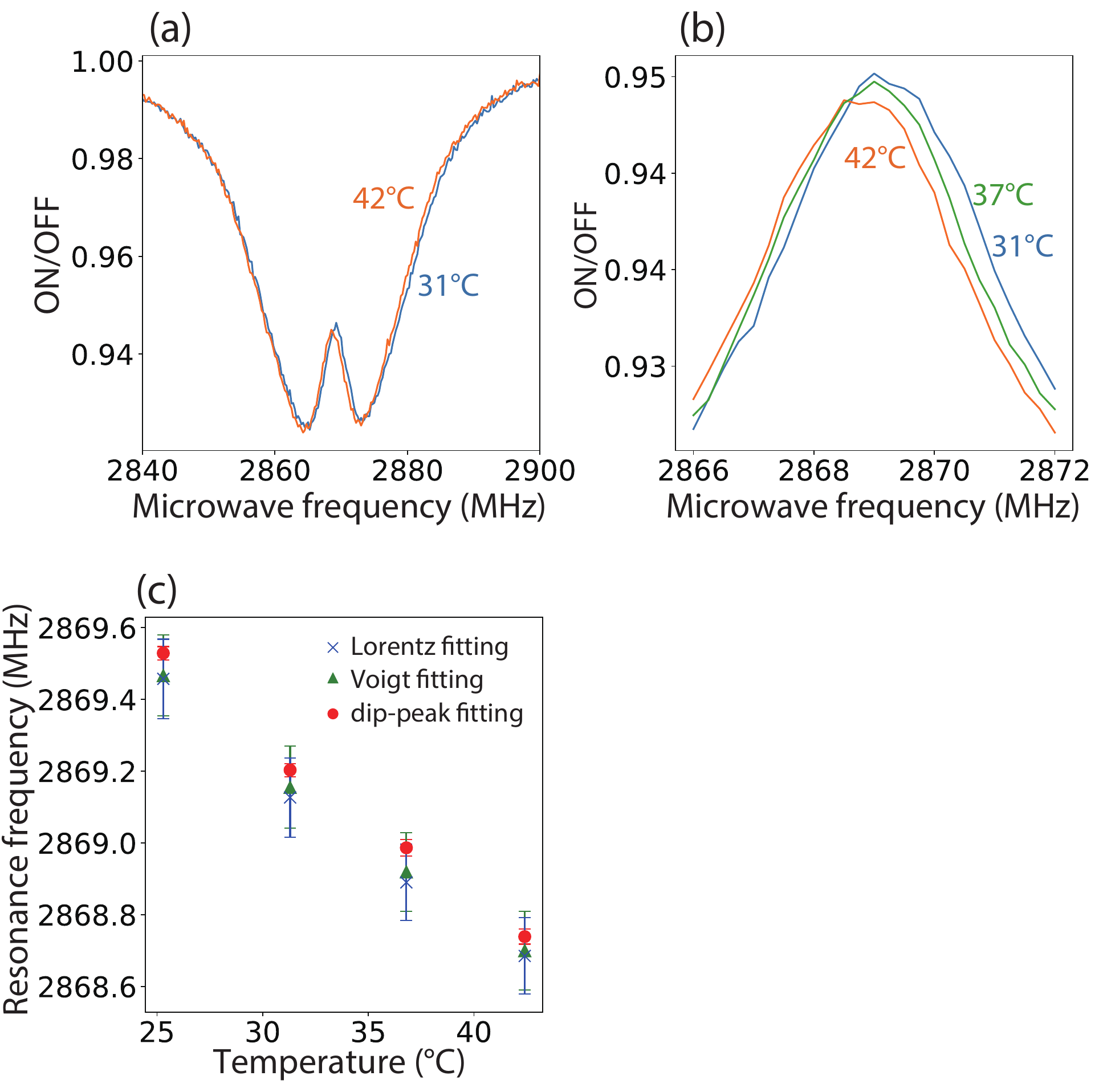}% Here is how to import EPS art
\caption{\label{fig:fig6}(a, b) Temperature dependence of the cw-ODMR spectra.
(c) Temperature dependence of the resonance frequency obtained by fitting the cw-ODMR spectra using a Lorentzian function (blue crosses), a Voigt function (green triangles), and the dip–peak fitting model (red circles).}
\end{figure}

We further investigated whether the resonance frequencies obtained using the dip–peak fitting exhibit the same temperature dependence as those obtained using conventional fitting methods when the temperature of FNDs was varied [Fig.~\ref{fig:fig6}]. The resonance frequency obtained using the dip–peak fitting shifts with temperature in a manner consistent with that obtained using conventional fitting approaches.

From these results, we conclude that the dip–peak fitting provides a reliable method for identifying the resonance frequency from cw-ODMR spectra for temperature-sensing applications. Compared with conventional fitting methods, the dip–peak fitting enables higher-precision determination of the resonance frequency within a shorter measurement time and operates effectively under lower microwave power conditions.

\section{\label{sec:level1}CONCLUSION}
In this study, we analytically clarified the physical origin of the dip–peak fitting method, a practical approach for modeling peak-like features observed within the dip near the resonance frequency in ensemble cw-ODMR spectra. By expressing the ensemble cw-ODMR spectrum as a convolution of single-NV spectra and transforming it analytically, we demonstrate that the spectral response around the resonance can be accurately described by a single Lorentzian function with a background term. This establishes a clear physical foundation for dip–peak fitting.

We further validate the effectiveness of this method experimentally. Compared with conventional fitting approaches, dip–peak fitting provides a more accurate description of the ODMR lineshape near resonance. Importantly, it enables more precise determination of the resonance frequency within a shorter measurement time, achieving an improvement in precision by approximately a factor of 1.6 under identical acquisition conditions. In addition, this method allows for more reliable estimation of confidence intervals based on fitting uncertainties, which has been difficult to achieve with conventional methods.

We find that dip–peak fitting maintains high accuracy over a wide range of microwave excitation strengths, with optimal performance achieved at lower microwave powers, corresponding to a contrast of approximately 5\%. Despite the reduced excitation strength, the temperature fluctuations measured using dip–peak fitting reach within a factor of ~10 of the shot-noise limit, whereas conventional methods remain at ~25 times the limit. Notably, both methods capture consistent temperature variations, confirming the reliability of dip–peak fitting.

Overall, dip–peak fitting provides a simple yet physically grounded and experimentally robust method for extracting resonance frequencies from cw-ODMR spectra. Its ability to achieve high precision with reduced microwave power and shorter acquisition time makes it particularly promising for high-accuracy temperature sensing using fluorescent nanodiamonds. This approach is expected to enable high-precision temperature measurements at the nanoscale.

\begin{acknowledgments}
We would like to acknowledge Dr. Hitoshi Ishida for his support in mathematics, and Dr. Hiroshi Abe and Dr. Takeshi Ohshima for the fabrication of FNDs. This work was supported by Early-Career Scientists (19K16089 and 21K15053 to S.S.), Toyota Riken Scholar Program (to S.S.), Kyoto Technoscience Center (to S.S.), JST-FOREST Program (JPMJFR2428 to S.S.), JSPS KAKENHI (22H02583 and 24H02306 to Y.H., and 25K08446 to K.F.), MEXT Q-LEAP (JPMXS0120330644 to Y.H.), and JST CREST (JPMJCR24B6 to Y.H.) 
\end{acknowledgments}

\appendix

\section{Monte Carlo simulations}
We perform Monte Carlo simulations of an ensemble of NV centers following the approach of Ref.~\cite{zhu_observation_2014, hayashi_optimization_2018}. 
The Hamiltonian of the NV center is described as
\begin{eqnarray}
H &=& D(T) S_z^2 + E_1 \left( S_x^2 - S_y^2 \right)\\
 &&+ E_2 \left( S_x S_y + S_y S_x \right)
 + \lambda \cos(\omega t)\, S_x
\end{eqnarray}
Applying the unitary transformation $U = e^{-i\omega S_z^2 t}$, the Hamiltonian in the rotating frame is written as
\begin{eqnarray}
    H' &=& (D - \omega) S_z^2+E_1\left(S_x^2-S _y^2\right)\\
    &&+E_2\left(S_x S_y+S_y S_x\right)+\frac{\lambda}{2} S_x\\
     &=& (D - \omega) ( \Ket{B}\Bra{B}+ \Ket{D}\Bra{D})\\
     &&+E_1( \Ket{B}\Bra{B}-\Ket{D}\Bra{D})\\
     &&+iE_2\left(\Ket{B}\Bra{D} - \Ket{D}\Bra{B}\right)+\frac{\lambda}{2} S_x
\end{eqnarray}
Where $\Ket{B}=(1/\sqrt{2})\left(\Ket{1}+\Ket{-1}\right)$, and $\Ket{D}=(1/\sqrt{2})\left(\Ket{1}-\Ket{-1}\right)$.
The time evolution of the density matrix $\rho$ under this Hamiltonian is analyzed using the Lindblad master equation to incorporate decoherence effects.
\begin{eqnarray}
    \frac{d\rho(t)}{dt}
&=& -\frac{i}{\hbar}\bigl[ H, \rho(t) \bigr]
+ \sum_{j=1}^{4} \gamma_{j}
[
2 L_{j} \rho(t) L_{j}^{\dagger}\\
&&- L_{j}^{\dagger} L_{j} \rho(t)
- \rho(t) L_{j}^{\dagger} L_{j}
].
\end{eqnarray}
where $L_1 = \Ket{B}\Bra{D},L_2 = \Ket{D}\Bra{B},L_3 = \Ket{B}\Bra{0}$, and $L_4 = \Ket{D}\Bra{0}$. The corresponding decay rates are taken as $\gamma_1 = \gamma_2 \equiv \gamma$, $\gamma_3 = \gamma_4 \equiv \gamma_{\text{rel}}$ 

\begin{figure}[t]
\includegraphics[width=\columnwidth]{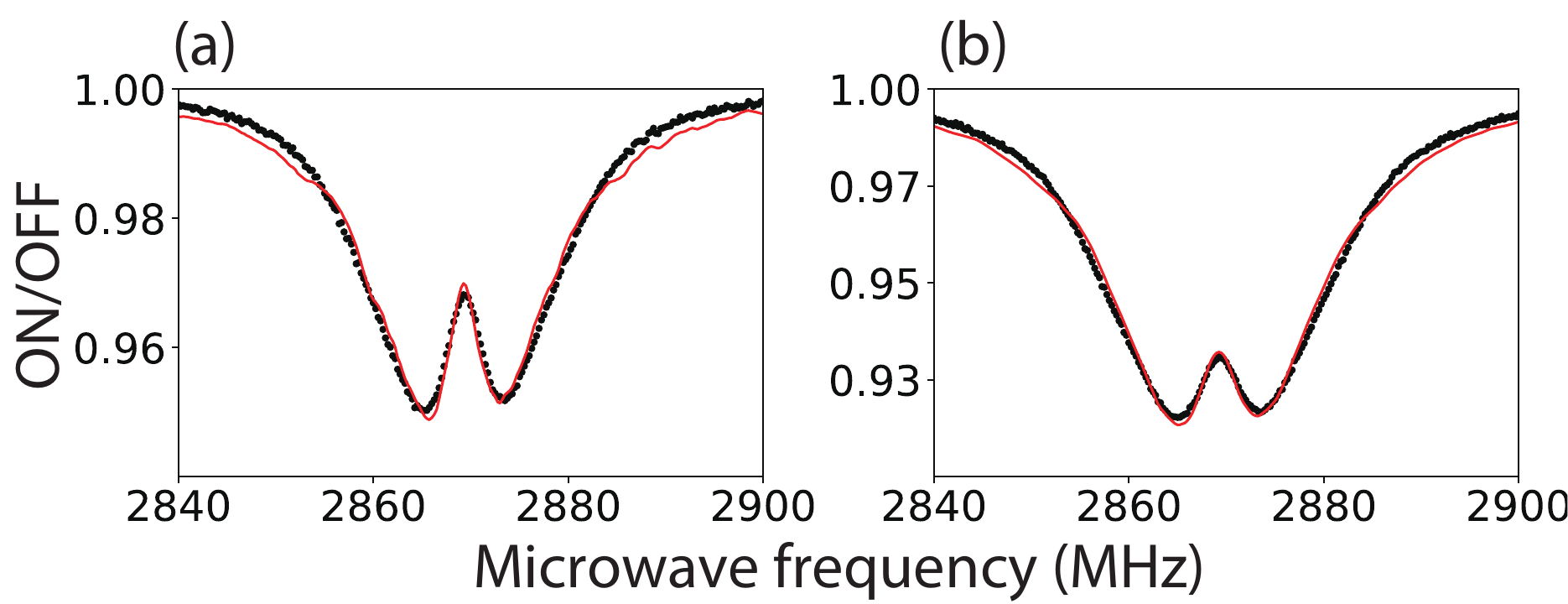}% Here is how to import EPS art
\caption{\label{fig:figS1}Monte Carlo simulation results of cw-ODMR spectra under weak microwave excitation (a) and strong microwave excitation (b). The simulation parameters were $\gamma = 0.076$ MHz, $\gamma_{\mathrm{rel}} = 0.0061$ MHz, $\eta = 0.57$, $\lambda = 0.41$ MHz, $D_0 = 2869.25$ MHz, $\gamma_D = 0.53$ MHz, $\gamma_{E_1} = 3.90$ MHz, $E_1^{0} = -0.34$ MHz, and $\gamma_{E_2} = 3.47$ MHz for (a), and $\gamma = 0.20$ MHz, $\gamma_{\mathrm{rel}} = 0.012$ MHz, $\eta = 0.59$, $\lambda = 0.74$ MHz, $D_0 = 2869.20$ MHz, $\gamma_D = 0.87$ MHz, $\gamma_{E_1} = 4.32$ MHz, $E_1^{0} = -0.21$ MHz, and $\gamma_{E_2} = 3.40$ MHz for (b).}
\end{figure}

By solving this equation, the cw-ODMR spectrum of a single NV center can be described. Furthermore, by solving the equation for a large number of NV centers and statistically averaging the results, the ensemble cw-ODMR spectrum can be obtained. In practice, this ensemble behavior is evaluated using Monte Carlo simulations. In the fitting procedure, the parameters listed in Table I were taken to be common to all NV centers, while the parameters listed in Table II were randomly varied for each NV center. Using this approach, the cw-ODMR spectra were fitted.

The results of this fitting are shown in Fig. 7. The cw-ODMR spectrum is well reproduced by this model, indicating that the essential spectral features are captured by the present description. These results suggest that the ensemble cw-ODMR spectrum arises from the superposition of sharp single-NV cw-ODMR spectra. Furthermore, the spectral variations of the individual single-NV cw-ODMR spectra are governed by fluctuations in the parameters $D$, $E_1$, $E_2$, which follow Lorentzian distributions.

\begin{table}[h]
\caption{\label{tab:table4}}
\begin{ruledtabular}
\begin{tabular}{ll}
Parameter&Physical meaning\\
\hline
$\gamma$ &Dissipation rate between the excited states \\&$\Ket{B}$ and $\Ket{D}$\\
$\gamma_{\mathrm{rel}}$ & Relaxation rate from the excited states $\Ket{B}$ or \\&$\Ket{D}$ to the ground state $\Ket{0}$.\\
$\eta$ &cw-ODMR detection efficiency, accounting for \\&the fact that fluorescence can be emitted \\&even when the system is in the excited-state.\\
$\lambda$ & Microwave drive strength (Rabi frequency)\\
\end{tabular}
\end{ruledtabular}
\end{table}

\begin{table}[h]
\caption{\label{tab:table4}}
\begin{ruledtabular}
\begin{tabular}{lll}
Parameter&Physical meaning&distribution\\
\hline
$D$ &Zero-field splitting& Lorentzian distribution with \\&&center $D_0$ and half width $\gamma_D$\\
$E_1$ &Lattice distortion& Lorentzian distribution with \\&&center $E_1^0 \simeq 0$ and half \\&&width $\gamma_{E_1}$\\
$E_2$ &Lattice distortion& Lorentzian distribution with \\&&center 0 and half width $\gamma_{E_2}$\\
\end{tabular}
\end{ruledtabular}
\end{table}
\newpage
\section{Visualization of the Lorentzian Approximation}
In the main text, we derive the analytical solution in Eq.~\eqref{eq:analytic_solution} and present an approximate form described by a single Lorentzian in Eq.~\eqref{eq:dip_peak}. Here, we verify the validity of this approximation by plotting both expressions.

To clarify the dependence of the spectral shape on the system parameters, we introduce the dimensionless variables
\begin{equation}
x=\frac{\omega-D_0}{\gamma_E},\qquad
g=\frac{\Gamma'}{\gamma_E},\qquad
\tilde{\lambda}=\frac{\lambda'}{\gamma_E},
\end{equation}
Eq.~\eqref{eq:analytic_solution} can be rewritten as

\begin{align}
P(x)=&1-\tilde{\lambda}\,\Biggl( L(x,0,\sqrt{2}+g) -L(x,0,\sqrt{2}-g)\\
&+\frac{8g}{\pi^2}
\Re\left[
\tilde{\beta}
\frac{1}{\sqrt{\tilde{u}_+}}
\ln\left(
\frac{1+\sqrt{\tilde{u}_+}}
{1-\sqrt{\tilde{u}_+}}
\right)
\right]
\Biggl)
\end{align}
where
\begin{equation}
\tilde{u}_+=(x+ig)^2+1,
\end{equation}
and
\begin{equation}
\tilde{\beta}
=
\frac{x+ig}
{2ig\left(x^2+2-g^2+2ixg\right)}.
\end{equation}
In this dimensionless representation, the spectral shape is
governed solely by the ratio \(g=\Gamma'/\gamma_E\),
apart from the overall amplitude factor
\(\tilde{\lambda}\).
To examine the validity of the Lorentzian approximation,
we compare the analytical expression with the approximate
Lorentzian form for several values of \(g\).
As shown in Fig.~8, the approximation accurately reproduces
the spectral feature near the resonance frequency \(D_0\)
over the range
\(0.2 < \Gamma'/\gamma_E < 0.6\).
\begin{figure}[h]
\includegraphics[width=\columnwidth]{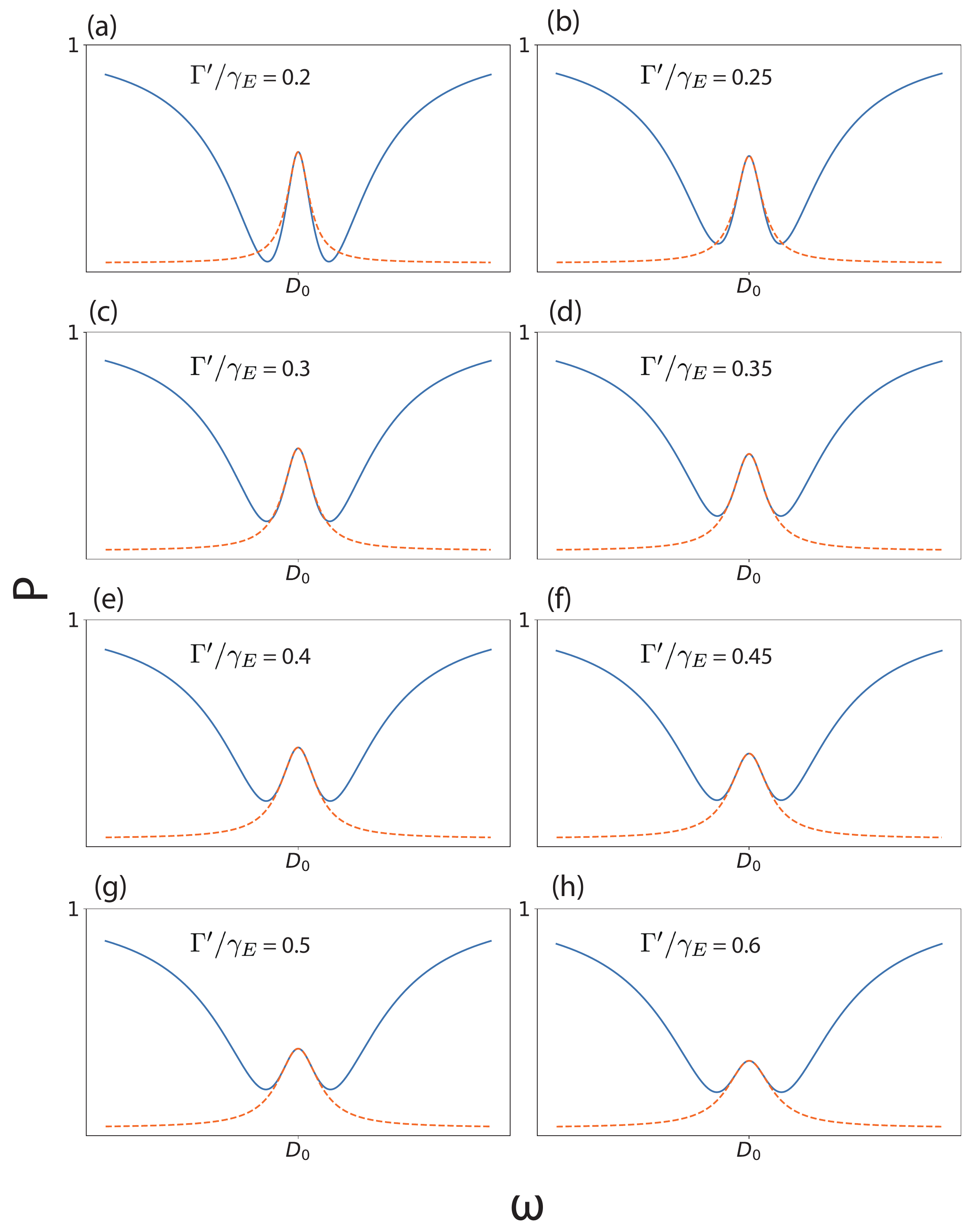}% Here is how to import EPS art
\caption{\label{fig:S2} Graphical representation of the analysis results for $\Gamma'/\gamma_E$ values of (a) 0.20, (b) 0.25, (c) 0.30, (d) 0.35, (e) 0.40, (f) 0.45, (g) 0.50, and (h) 0.60. The blue solid line represents P in eq 8, and the orange dashed line represents $A + B\,L(\omega,D_0,\Gamma_{\mathrm{dip}})$ in eq 11.
}
\end{figure}
\newpage

\section{Application of Dip–Peak Fitting to Various FNDs}
\begin{figure*}[t]
\includegraphics[width=\textwidth]{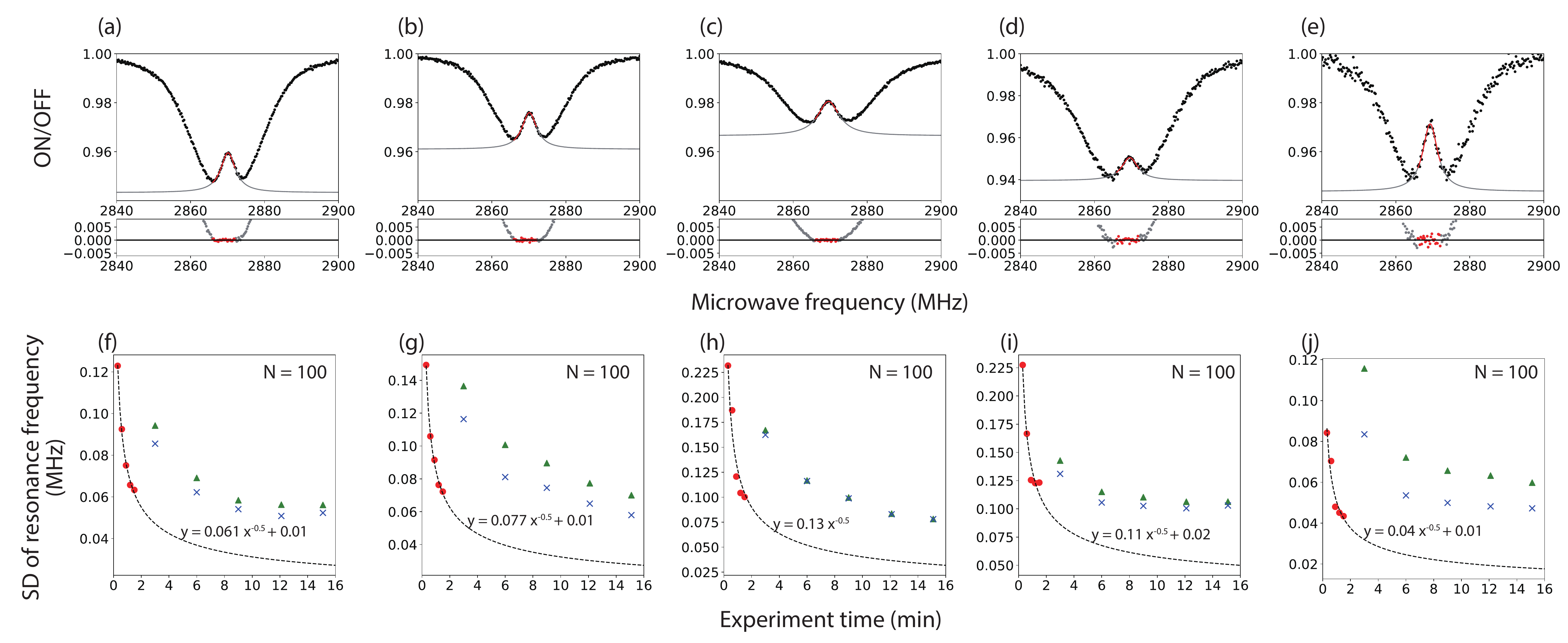}% Here is how to import EPS art
\caption{\label{fig:S3} Results of dip–peak fitting for 100 nm FNDs with electron irradiation doses of $4 \times 10^{18}~\mathrm{cm^{-2}}$ (a, f), $6 \times 10^{18}~\mathrm{cm^{-2}}$ (b, g), and $1 \times 10^{19}~\mathrm{cm^{-2}}$ (c, h), as well as for commercially available FNDs of 100 nm (d, i) and 40 nm (e, j). (a, b, c, d, e) Black dots represent the experimental cw-ODMR data. The fitting results are shown as red solid lines within the fitting range and as gray solid lines outside the range. Residuals shown below each spectrum correspond to the differences between the experimental data and the fitting curves; those within the fitting range are shown as red dots, while those outside the range are shown as gray dots. (f, g, h, i, j)
Dependence of the resonance-frequency fluctuation on the measurement time, obtained from fitting ensemble cw-ODMR spectra composed of 100 bright spots. Blue crosses indicate results obtained using Lorentzian fitting, green triangles correspond to Voigt fitting, and red circles represent results obtained using the dip–peak fitting model.
The dashed lines represent fits to the resonance-frequency fluctuations obtained using the dip–peak fitting model, assuming a scaling proportional to the inverse square root of the measurement time. The corresponding fitting parameters are indicated in the figure.
}
\end{figure*}
To evaluate the applicability of the dip–peak fitting across different FNDs, we investigated its performance not only for the 100 nm FND with an electron irradiation dose of 
$4 \times 10^{18} \text{cm}^{-2}$
, which is primarily used in this study, but also for FNDs with different NV center concentrations corresponding to irradiation doses of $6 \times 10^{18} \text{cm}^{-2}$ and $10 \times 10^{18} \text{cm}^{-2}$. 
In addition, commercially available 100 nm and 40 nm FNDs were examined [Fig. 9].
For all tested samples, dip–peak fitting consistently enabled faster and more accurate determination of the resonance frequency than conventional methods for cw-ODMR spectra with a contrast of approximately 5\%. These results demonstrate the robustness and broad applicability of the dip–peak fitting approach.

\newpage

\bibliography{references2}% Produces the bibliography via BibTeX.

\end{document}